\documentclass[11pt]{article}
\usepackage{latexsym}
\usepackage{graphicx, epstopdf}
\usepackage{amsfonts}
\usepackage{amsmath, amsthm, amssymb}
\usepackage{fullpage}
\usepackage{setspace}
\usepackage{longtable}
\usepackage{natbib}
\bibpunct{(}{)}{;}{a}{}{,}
\usepackage{dcolumn}
\newcolumntype{.}{D{.}{.}{-1}}
\newcolumntype{d}[1]{D{.}{.}{#1}}
\usepackage[top=.8in, left=.5in, right=.5in,
bottom=.8in]{geometry}
\usepackage{bm}
\usepackage[compact]{titlesec}
\usepackage{booktabs}
\usepackage{enumitem}

\usepackage{color}
\RequirePackage[colorlinks,citecolor=blue,urlcolor=blue]{hyperref}
\usepackage{amsthm}

\newtheorem{theorem}{Theorem}
\newtheorem{lemma}{Lemma}
\newtheorem{remark}{Remark}

\usepackage{caption}
\usepackage{subcaption}
\usepackage[linesnumbered, ruled, vlined]{algorithm2e}

\expandafter\def\expandafter\normalsize\expandafter{\normalsize\setlength\abovedisplayskip{0pt}}
\expandafter\def\expandafter\normalsize\expandafter{\normalsize\setlength\belowdisplayskip{0pt}}
\expandafter\def\expandafter\normalsize\expandafter{\normalsize\setlength\abovedisplayshortskip{0pt}}
\expandafter\def\expandafter\normalsize\expandafter{\normalsize\setlength\abovedisplayshortskip{0pt}}

\newcommand*\samethanks[1][\value{footnote}]{\footnotemark[#1]}

\begin{document}
\pagestyle{plain}

\title{Estimation and inference on high-dimensional individualized treatment rule in observational data using split-and-pooled de-correlated score}

\author{
  Muxuan Liang \thanks{Public Health Sciences Divisions, Fred Hutchinson Cancer Research Center}\\
  \and
  Young-Geun Choi \thanks{Department of Statistics, Sookmyung Women's University}\\
  \and
  Yang Ning \thanks{Department of Statistics and Data Science, Cornell University}\\
  \and
  Maureen A Smith \thanks{Departments of Population Health and Family Medicine, University of Wisconsin-Madison}\\
  \and 
  Ying-Qi Zhao \samethanks[1]\\
}
\maketitle
\thispagestyle{empty}

\abstract{
With the increasing adoption of  electronic health records,  there is an increasing interest in developing individualized treatment rules, which recommend treatments according to patients' characteristics, from large observational data. However,  there is a lack of valid inference procedures for such rules developed from this type of data in the presence of high-dimensional covariates. In this work, we develop a penalized doubly robust method to estimate the optimal individualized treatment rule from high-dimensional data.  We propose a split-and-pooled de-correlated score to construct hypothesis tests and confidence intervals. Our proposal utilizes the data splitting to conquer the slow convergence rate of nuisance parameter estimations, such as non-parametric methods for outcome regression or propensity models. We establish the limiting distributions of the split-and-pooled de-correlated score test and the corresponding one-step estimator in high-dimensional setting. Simulation and real data analysis are conducted to demonstrate the superiority of the proposed method.
}

\newcommand{\n}{\noindent}
{\bf Keywords:} 	Individualized treatment rule; double-robustness; high-dimensional inference; semiparametric inference; precision medicine. 

\newpage

\section{Introduction}
	
	An individualized treatment rule is a decision rule that maps the patient profiles $X\in \mathcal{X}$, a subspace of $R^p$, into the intervention space $A\in\mathcal{A}$, where $p$ is the number of the covariates and $\mathcal{A}$ is the set of available interventions. Given an outcome of interest, the optimal individualized treatment rule maximizes the value function which is the mean outcome if it were applied to a target population. Understanding the driving factors of a data-driven treatment rule can help with identifying the source of the heterogeneous effects and with guiding practical applications of precision medicine.   
	
	The increasing adoption of electronic health records at healthcare centers has provided us unprecedented opportunities to understand the optimal individualized treatment rule through massive observational data. One of the difficulties in dealing with observational data is the high dimensionality of the covariates. There have been various methods developed to estimate the optimal individualized treatment rule.	For regression-based approaches, Q-learning methods \citep{Watkins1992, chakraborty2010inference,qian2011, Laber2014} pose a fully specified model assumption on the conditional mean of the outcomes given the covariates and treatments. \citet{qian2011} approximates the conditional mean by a rich linear model, along with an $l_1$ penalty to accommodate high-dimensional data.  A-learning methods \citep{Murphy2003optimal, Wenbin2011, shi2016, shi2018} pose a model assumption on the contrast function of the conditional means. With high-dimensional covariates, \citet{shi2016, shi2018} adopt penalized estimating equation or penalized regression with a linear contrast function.  An alternative class of methods searches over a pre-specified class of individualized treatment rules to optimize an estimator of the mean outcome, usually called direct \citep[][]{laber2014dynamic}, policy learning \citep{athey2017efficient} or value-search \citep[][]{marieChapter} estimators. Among these methods, \citet{yingqi2012} propose the outcome weighted learning approach based on an inverse probability weighted estimator of the value.
	\citet{song2015sparse} develop a variable selection method based on penalized outcome weighted learning for optimal individualized treatment selection. 
	
	Statistical inference for the optimal or estimated individualized treatment rule is particularly challenging in the presence of high-dimensional covariates. Confounding and selection bias presented in large observational data such as EHR data add one more layer of complexity.  \citet{shuhan2018} propose a concordance-assisted learning algorithm in the presence of high-dimensional covariates. Nonetheless, they do not provide any inference procedures. 
Inference methods for A-learning approaches such as \citet{song2017} and \citet{jeng2018} are developed assuming the  propensity score is known. Thus, their methods cannot be applied if data are collected from observational studies. \citet{shi2018} derive the oracle inequalities of the proposed estimators for the parameters in a linear contrast function,  but their work focuses on the selection consistency and has little discussion on the inference of the estimated rule. Their method depends on parametric assumptions on the propensity and outcome models, and thus may not be consistent when complex propensity or outcome models are expected. In practice, to avoid misspecification, flexible models may be adopted for the outcome regression or the propensity score. However, these models result in slow convergence rates for the nuisance parameters, and deteriorate the limiting distribution of the estimated decision rule. As such, it is important to propose an inference procedure for the estimated decision rule, which is valid under the high-dimensional setup and robust to flexible models for the nuisance parameters. Recent literature on the high-dimensional inference can assist with tackling this challenge. For example,  \citet{vandegeer2014} propose a debiased Lasso approach for generalized linear models. \citet{ning2017}  propose a de-correlated score test for low dimensional parameters with the existence of the high-dimensional covariates, which is applicable for parametric models with correctly specified likelihoods. \citet{chunhui2017} propose a bootstrap procedure for high-dimensional inference, but it is computationally intensive. 

Another importance and related topic is the inference of the optimal value. The inference of the optimal value has been shown to be challenging at exceptional laws (non-regular case) where there exists a subgroup of patients for which treatment effect vanishes \citep{chakraborty2010inference, laber2014rejoinder, goldberg2014comment}. To achieve the inference of the optimal value in low-dimensional setup, \cite{chakraborty2014inference} propose an $m\text{-out of-}n$ bootstrap to construct a confidence interval for the value. \cite{luedtke2016statistical} propose an online one-step estimator which is the weighted  average the values estimated on chunks of data increasing in size. Recently, \cite{shi2020breaking} use a subagging algorithm to aggregate value estimates obtained by repeated sample splittings. In both \cite{luedtke2016statistical} and  \cite{shi2020breaking}, a single-split procedure is also discussed to facilitate the computation, though the resulting confidence interval might be wider. However, the value inference for high-dimensional setup is lacking.
	
    In this work, we propose a novel penalized doubly robust approach, termed as penalized efficient augmentation and relaxation learning, to estimate the optimal individualized treatment rule in observational studies with high-dimensional covariates.  We construct the decision rules by optimizing a convex relaxation of the augmented inverse probability weighted estimator of the value with penalties, which generalizes the method proposed in \citet{zhao2019efficient} to high-dimensional setup. The proposed procedure involves estimation of the conditional means of the outcomes and the propensity scores as nuisance parameters. As long as one of the nuisance models is correctly specified, we can consistently estimate the optimal individualized treatment rules under certain conditions. 
 Furthermore, we propose a split-and-pooled de-correlated score test, which provides valid hypothesis testing and interval estimation procedures to identify the driving factors of the estimated decision rule. The proposed procedure generalizes the de-correlated score \citep{ning2017} to handle the potential slow convergence rates from the  nuisance parameters estimation and to allow a general loss function. Sample-splitting is adopted to separate the estimation of the nuisance parameters from the construction of the de-correlated score,  which is utilized in \citet{victor2018} for inference on a low-dimensional parameter of interest in the presence of high-dimensional nuisance parameters. However, the inference on the estimated decision rule using the proposed approach requires a more sophisticated analysis due to the convex relaxation schemes. Theoretically, we show that the split-and-pooled de-correlated score is asymptotically normal even when the nuisance parameters are estimated non-parametrically with slow convergence rates. In addition, we use a single-split procedure to infer the value under the estimated decision rule.
	
	\section{Method}
	\label{sec:method}
	
	\subsection{Penalized efficient augmentation and relaxation learning } 
	\label{sec:earl}
	
	 Let $X$ be a $p$-dimensional random vector, which contains the baseline covariates capturing patient profiles. We assume that $p$ can be much larger than the sample size $n$. Let $A\in\{-1, 1\}$ be the treatment assignment, and $Y\in R$ be the observed outcome that higher values are preferred. Here, we adopt the framework of potential outcomes \citep{rubin1974, rubin2005}. Denote the potential outcome under treatment $a\in\{-1, 1\}$ as $Y(a)$. Then the observed outcome is $Y=Y(a)I\{a=A\}$, where $I\{\cdot\}$ is the indicator function. An individualized treatment rule, denoted by $D$,  is a mapping from the space of covariates $\mathcal{X} \subseteq {R}^p$ to the space of treatments $\mathcal{A}=\{-1, 1\}$. With a slight abuse of notation, we write the observed outcome under this decision rule as $Y(D)=\sum_{a\in\{-1,1\}} Y(a)I\{a=D(X)\}$. The expectation of $Y(D)$, $V(D) = {E}\left(Y(D)\right)$, is called the \textsl{value function} which is the average of the outcomes over the population if the decision rule were to be adopted. In order to express the value in terms of the data generative model, we assume the following conditions: 1) the stable unit treatment value assumption \citep{imbens_rubin_2015}; 2) the strong ignorability $Y(-1),Y(1)\perp A\mid X$; 3) Consistency $Y = Y(A)$. The stable unit treatment value assumption assumes that the potential outcomes for a patient do not vary with the treatments assigned to other patients. It also implies that there are no different versions of the treatment. The strong ignorability condition means that there is no unmeasured confounding between the potential outcomes and the treatment assignment mechanism. The optimal individualized treatment rule is defined as $D_{\mathrm{opt}} = \arg\max_D\{V(D)\}$.  
	 
	In this paper, due to the high-dimensional nature of the data we work with, we focus on deriving a linear decision rule of the form $D(x) = \mathrm{sgn}(x^\top\beta)$, where $x\in\mathcal{X}$ and the function $\mathrm{sgn}(t)=1$ if $t\geq 0$; $\mathrm{sgn}(t)=-1$ if $t<0$. In general, $D_{\mathrm{opt}}(x)$ could be a complex function of $x$, but in many situations, the optimal rule $D_{\mathrm{opt}}(x)$ may only depend on a linear function of $x$ \citep{xu2015regularized}.  We assume that $D_{\mathrm{opt}}(x) = \mathrm{sgn}(x^\top \beta^{\mathrm{opt}})$, which also indicates $D_{\mathrm{opt}}( x) = \mathrm{sgn}(cx^\top\beta^{\mathrm{opt}})$ for any $c > 0$. To avoid $\beta^{\mathrm{opt}}=0$ and identifiability issue, we will restrict estimation to regimes in which $\|\beta^{\mathrm{opt}}\|_2>1$.
	
Let $\pi(a;x)=\mathrm{pr}(A=a\mid X=x)$ and $Q(a;x)= E (Y\mid X=x, A = a)$ for $a\in \{-1, 1\}$ and $x\in R^p$. Define the weights
	\begin{equation*}
	\widehat{W}_a = W_a(Y,X,A,\widehat\pi,\widehat Q) = \frac{YI\left\{A=a\right\}}{\widehat{\pi}(a;X)}- \frac{\left [I\left\{A=a\right\}-\widehat{\pi}(a;X)\right]\widehat{Q}(a;X)}{\widehat{\pi}(a;X)}
	\end{equation*}
	for $a\in\{-1,1\}$, where $\widehat{\pi}(a;  X)$ and $\widehat{Q}(a;  X)$ are the estimators of $\pi(a;  X)$ and $Q(a;  X)$ respectively. Under the conditions above, the augmented inverse probability weighted estimator of the value function is
	\begin{equation*}
	\widehat{V}(D)={E}_n\left[\widehat{W}_1I\left\{D(X)=1\right\}+\widehat{W}_{-1}I\left\{D(X)=-1\right\}\right],
	\end{equation*}
	where ${E}_n[\cdot]$ denotes the empirical average. The estimator $\widehat{V}(D)$ enjoys the double robustness property. Assume that $\widehat{Q}(a; x)$ and $\widehat{\pi}(a;x)$ converge in probability uniformly to some deterministic limits, denoted by $Q^{m}(a; x)$  and $\pi^{m}(a;x)$, respectively. $\widehat{V}(D)$ converges to $V^m(D)$, where
	\begin{equation*} 
	 {V}^m(D)= {E}\left[{W}_{1}^m I\left\{D(X) = 1\right\}+{W}_{-1}^m I\left\{D(X) = -1\right\}\right]. 
	\end{equation*}
Here,  ${W}_a^m = W_a(Y, X,A, \pi^m, Q^m)$ is the limit that $\widehat{W}_a$ converges to, $a=\pm 1$. 
As shown in \citet{zhao2019efficient}, if  either ${\pi}^m(a;  x) = \pi(a;  x)$ or ${Q}^m(a;  x) = Q(a;  x)$, but not necessarily  both, then $V^m(D) = {V}(D)$.
	
To avoid negative $\widehat{W}_a$, we consider its positive and negative parts separately and define $\widehat{W}_{a,+} = |\widehat{W}_{a} |1 \{\widehat{W}_{a}\geq 0\}$ and $\widehat{W}_{a,-} = |\widehat{W}_{a} |1 \{\widehat{W}_{a}\leq  0\}$. Maximizing $\widehat{V}(D)$ is equivalent to minimizing 
\begin{equation}\label{eq:emp_loss}
 {E}_n\left[\left(\widehat{W}_{1,+}+ \widehat{W}_{-1,-}\right)I\left\{D(X)\neq 1\right\}+\left(\widehat{W}_{1,-}+ \widehat{W}_{-1,+}\right)I\left\{D(X) \neq -1\right\}\right]. 
\end{equation}
Directly optimizing \eqref{eq:emp_loss} is infeasible due to the indicator functions  in the objective function, especially with a large number of covariates. To avoid minimizing the indicator function, we replace the indicator function with a strictly convex surrogate loss. Due to the strict convexity, the minimizer of the surrogate loss is always unique. Thus, we can relax the constraint that $\|\beta\|_2> 1$. Furthermore, we add a sparse penalty function, which enables us to eliminate the unimportant variables from the derived rule. We denote the weight encouraging $A=1$ as $\widehat{\Omega}_+=\widehat{W}_{1,+}+\widehat{W}_{-1,-}$ and the weight encouraging $A=-1$ as $\widehat{\Omega}_-=\widehat{W}_{1,-}+ \widehat{W}_{-1,+}$. Our proposed estimator $\widehat{\beta}$ is  
	\begin{equation}\label{eq:loss_empirical}
	\widehat{\beta} = \arg\min_{\beta}E_n\left[\widehat{\Omega}_+\phi\left (X^\top\beta\right )+\widehat{\Omega}_-\phi\left (- X^\top\beta\right ) \right]+\lambda_nP(\beta),
	\end{equation}
	where $\phi$ is a convex surrogate loss, $P(\beta)$ is a sparse penalty function with respect to $\beta$, and $\lambda_n$ is a tuning parameter controlling the amount of penalization. In this paper, we focus on the $L_1$ lasso penalty $P(\beta)=\|\beta\|_1$. The framework allows a broad class of surrogate loss functions, such as logistic loss, $\phi(t)=\log\left(1+e^{-t}\right)$, see Section \ref{sec:theoretical} for the detailed technical conditions on $\phi$. The estimated decision rule can be subsequently obtained as $\widehat{D}(X)=\mathrm{sgn}\left(X^\top\widehat{\beta}\right)$.
	
	\subsection{Split-and-pooled de-correlated score test}
	\label{sec:inference}
        We define
	\begin{equation*}
	l_\phi(\beta;\Omega_+^m,\Omega_-^m)=\Omega_+^m\phi\left (X^\top\beta\right)+\Omega_-^m\phi\left(-X^\top\beta\right),
	\end{equation*}
and  ${\beta^*} = \arg\min_{\beta}{E}\left[l_\phi(\bm\beta;\Omega_+^m,\Omega_-^m)\right],$ where ${\Omega}_+^m=  {W}_{1,+}^m + {W}_{-1,-}^m$ and ${\Omega}_-^m= {W}_{1,-}^m + {W}_{-1,+}^m$. To simplify notations, we will suppress the superscript and write them as $\Omega_+$ and $\Omega_-$ instead.  Let $X=(X_1, X_{-1})$ where $X_1\in {R}$ is the first covariate and $X_{-1}\in {R}^{p-1}$ includes the remaining covariates. Likewise, let $\beta_1^*$ be the first coordinate of $\beta^*$ and ${\beta}_{-1}^*$ be a $p-1$ dimensional sub-vector of ${\beta}^*$ without $\beta_1^*$. Without loss of generality, suppose that $\beta_1^*$ is of interest. The statistical inferential problem can be formulated as testing the null hypothesis
$
	H_0: \beta_1^*=0\text{ versus     }H_1: \beta_1^*\not = 0,
$
or constructing confidence intervals for $\beta_1^*$. The proposed method can be easily generalized to the setting where $\beta_1^*$ is multi-dimensional. 
 	
Before we propose our inference procedure for $\beta^*$, we introduce a lemma to show that under certain conditions, our inference procedure for $\beta^*$ can provide information on $\beta^{\rm opt}$. Specifically, lemma~\ref{lemma:consistency} provides sufficient conditions that $\beta^*$ satisfies $D_{\mathrm{opt}}(X)=\mathrm{sgn}(X^\top\beta^{\mathrm{opt}})=\mathrm{sgn}(X^\top\beta^*)$.  
	
Define two subspaces depending on $\beta$,
 \begin{eqnarray*}
 	\Delta_{\phi}(\beta)&=&\left\{f(X)\in L_2: \mathrm{cov}\left[f(X),\left\{\phi(X^\top\beta)-\phi(-X^\top\beta)\right\}\mid X^\top\beta^{\rm opt}\right]\geq 0\right\},\\
 	S_{\phi}(\beta)&=&\left\{f(X)\in L_2: \mathrm{cov}\left[f(X),\left\{\phi(X^\top\beta)+\phi(-X^\top\beta)\right\}\mid X^\top\beta^{\rm opt}\right]\geq 0\right\}.
 \end{eqnarray*}
\begin{lemma}\label{lemma:consistency}
If the $D_{\mathrm{opt}}(X)$ has a linear form, and $Q^m = Q$ or $\pi^m = \pi$ in $\Omega_+$ and $\Omega_-$, then $D_{\mathrm{opt}} (X)=\mathrm{sgn}(X^\top\beta^*)$ if the following conditions are satisfied: (a) The contrast function ${E}(Y(1)-Y(-1)\mid X)\in \Delta_{\phi}(\beta^*)$, and the main effect ${E}(Y(1)+Y(-1)\mid X)\in S_{\phi}(\beta^*)$; (b) there exists a $p$-dimensional vector $P$ such that ${E}(X\mid X^\top \beta^{\mathrm{opt}})=P X^\top \beta^{\mathrm{opt}}$. 
\end{lemma} 

The subspaces $\Delta_{\phi}(\beta)$ and $S_{\phi}(\beta)$ enjoy the following properties: (i) Any measurable function of $ X^\top\beta_{\rm opt}$ belongs to $\Delta_{\phi}(\beta)\cap S_{\phi}(\beta)$, $\forall \beta$; (ii) Suppose that a function $g(X)\in \Delta_{\phi}(\beta)$ (or $S_{\phi}(\beta)$), then the function $h(X^\top\beta^{\rm opt})g(X)\in \Delta_{\phi}(\beta)$ (or $S_{\phi}(\beta)$), where $h(\cdot)$ is an arbitrary measurable function. Thus, if $E(Y_{1}\mid X)$ and $E(Y_{-1}\mid X)$ only depend on $ X^\top\beta^{\rm opt}$, Condition~(a) is easily satisfied. We provide  examples  in the supplementary materials (see pages~8-10) to further show that Condition (a)  is satisfied by a large class of models, including data generative models that are not single index models. 

Condition (b) on the design matrix $X$ is common in the dimension reduction literature \citep{li1991sliced, zhu2006sliced, Lin2018, Lin2019}. It is satisfied if the distribution of $X$ is elliptically symmetric. \citet{li1989regression, duan1991slicing} provide a thorough discussion on this condition in regression methods which aims to estimate a single index with an arbitrary and unknown link function. More specifically, they provide a bias bound when the elliptical symmetry is violated and show that the asymptotic bias is small when the elliptical symmetry is nearly satisfied. Further, \citet{hall1993almost} shows that when the dimension of $X$ is large, for most directions $\beta^{\rm opt}$ even the most nonlinear regression is still nearly linear. In addition, empirical studies by Brillinger and others suggest that quite often the bias may be negligible even for a moderate violation of condition (b) \citep{brillinger2012generalized, li1989regression}.

\begin{remark}
\noindent Alternatively, instead of assuming the conditions in Lemma~\ref{lemma:consistency}, the desired relationship $D_{\mathrm{opt}}(X)=\mathrm{sgn}(X^\top\beta^*)$  may still hold under some parametric assumptions on ${E}(Y_1\mid X)$ and ${E}(Y_{-1}\mid X)$. For example, if the outcomes are non-negative and the following conditions are satisfied
 \begin{equation}\label{cond:parametric}
\log\left\{E(Y_1\mid X)/E(Y_{-1}\mid X)\right\}=X^\top\beta^{\rm opt},
\end{equation}
we still have $D_{\mathrm{opt}}(X)=\mathrm{sgn}(X^\top\beta^*)$. Condition (\ref{cond:parametric}) poses a parametric assumption on ${E}(Y_1\mid X)/{E}(Y_{-1}\mid X)$ ({see supplementary material for the details}).   This ratio measures the relative change of the potential outcomes. Under Condition~(\ref{cond:parametric}),  hypothesis testing of $\beta^*$ is equivalent to testing for the driving factors of the $D_{\mathrm{opt}}$. Furthermore, the interval estimation of $\beta^*$ can be interpreted through the specified model assumption in~(\ref{cond:parametric}).
\end{remark}

Next, we introduce our proposed inference procedure. Suppose that $\Omega_+$ and $\Omega_-$ are known, then the estimator $\widehat{\beta}$ is obtained by minimizing the empirical loss
$
	{E}_n\left[l_\phi(\beta;\Omega_+,\Omega_-)\right]+\lambda_nP(\beta).
$
The score function of $\beta_1$ is
$
\mathrm{E}_n\left[\nabla l_\phi(\bm\beta;\Omega_+,\Omega_-)X_1\right],
$ 
where  $\nabla l_\phi(\beta;\Omega_+,\Omega_-)=\Omega_+\phi'\left(X^\top\beta\right)-\Omega_-\phi' \left (- X^\top\beta\right)$. Let $\widehat{\beta}_{\rm null}^\top=\left(0,\widehat{\beta}_{-1}^\top\right)$, where $\widehat{\beta}_{-1}$ is a $p-1$ dimensional sub-vector of $\widehat{\beta}$ without $\widehat{\beta}_1$. In the low dimensional setting where $p$ is fixed, the score function with $\widehat{\beta}_{\rm null}$, ${E}_n\left[\nabla l_\phi(\widehat{\beta}_{\rm null};\Omega_+,\Omega_-)X_1\right]$, is asymptotically normal. Nevertheless, in a high-dimensional setting, the asymptotic normality of the score function ${E}_n\left[\nabla l_\phi(\widehat{\beta}_{\rm null};\Omega_+,\Omega_-)X_1\right]$ is deteriorated by the high dimensionality of $\widehat{\beta}_{-1}$. Following \citet{ning2017}, we utilize the semiparametric theory to de-couple the estimation error of $\widehat{\beta}_{-1}$ with the score function of $\beta_1$.  A de-correlated score function is defined as 
$
	{E}_n\left[\nabla l_{\phi}(\widehat{\beta}_{\rm null}; \Omega_+, \Omega_-)\left(X_1-X_{-1}^\top w^*\right)\right],
$
where $ w^*=\left( I_{-1,-1}^*\right)^{-1} I_{-1,1}^*$ is chosen to reduce the uncertainty of the score function due to the estimation error of $\widehat{\beta}_{-1}$, and $ I_{-1,-1}^*$ and $ I_{-1,1}^*$ are the corresponding partitions of $ I^*={E}\left[\nabla^2 l_{\phi}(\beta^*; \Omega_+, \Omega_-) X X^\top\right]$.
	
Under the null hypothesis, this de-correlated score function follows
	\begin{equation*}
	n^{1/2}{E}_n\left[\nabla l_{\phi}(\widehat{\beta}_{\rm null}; \Omega_+, \Omega_-)\left(X_1- X_{-1}^\top w^*\right)\right]\to N\left(0, \left(\nu^*\right)^\top E\left[\nabla^2 l_{\phi}(\beta^*; \Omega_+, \Omega_-)XX^\top\right]\nu^*\right ),
	\end{equation*}
	where
	$\nabla^2 l_\phi(\beta;\Omega_+,\Omega_-)=\Omega_+\phi^{''}\left(X^\top\beta\right)+\Omega_-\phi^{''}\left(- X^\top\beta\right)$, and $\left(\nu^*\right)^\top=\left(1, -\left( w^*\right)^\top\right)$.  We propose to estimate the nuisance parameter $w^*$  via
	\begin{equation*}
	\min_{ w} {E}_n\left[\nabla^2 l_\phi\left(\widehat{\beta};{\Omega}_+,{\Omega}_-\right)\left(X_1- X_{-1}^\top w\right)^2\right] + \widetilde{\lambda}_{n}\| w\|_1,
	\end{equation*}
	where $\widetilde{\lambda}_n$ is a tuning parameter. Denote the estimator for $w^*$ as $\widehat{w}$. A valid test for $H_0: \beta_1^*=0$ is constructed based on 
	\begin{equation}\label{eq:de-correlated_score}
	{E}_n\left[\nabla l_{\phi}(\widehat{\beta}_{\rm null}; \Omega_+, \Omega_-)\left(X_1- X_{-1}^\top \widehat{w}\right)\right].
	\end{equation}

	The nuisance parameters, $\Omega_+$ and $\Omega_-$  are unknown in practice, and are estimated via modeling $\pi$ and $Q$. To avoid misspecification, they can be estimated using   flexible nonparametric or machine learning methods, which  may lead to convergence rates slower than $n^{-1/2}$. To overcome the possible slow convergence rates of $\widehat{\pi}$ and $\widehat{Q}$, we propose a split-and-pooled de-correlated score, where we consider a sample split procedure in constructing the de-correlated score function \citep{victor2018}. 
	
	Let ${I}_1, \ldots, {I}_K$ be a random partition of the observed data with approximately equal sizes, where $K\geq 2$ is a fixed pre-specified integer. We assume that $\lfloor n/K\rfloor\leq |{I}_k| \leq \lfloor n/K\rfloor+1$, for all $k=1,\ldots, K$.  Let ${E}_n^{(k)}[\cdot]$ denote the expectation defined by the data in ${I}_k$.  For each $k\in \{1,\ldots, K\}$, we repeat the following procedure. First, we  obtain $\widehat{\pi}_{(-k)}$ and $\widehat{Q}_{(-k)}$ using the data excluding ${I}_k$. In the presence of high-dimensional covariates, we can use generalized linear model with penalties \citep{vandegeer2008} or kernel regression after a model-free variable screening \citep{runze2012, Hengjian2015} for estimating $\pi$ and $Q$. A data-split estimator $\widehat{\beta}^{(k)}$ is obtained by 
\begin{equation}\label{eq:est_beta}
	\widehat{\beta}^{(k)} = \arg\min_{\beta}{E}_n^{(k)}\left[l_\phi\left(\beta;\widehat{\Omega}_+^{(-k)} ,\widehat{\Omega}_-^{(-k)}\right)\right]+\lambda_{n,k}\|\beta\|_1,
\end{equation}
	where $\widehat{\Omega}_+^{(-k)}$ and $\widehat{\Omega}_-^{(-k)}$ are  computed with $\widehat{\pi}_{(-k)}$ and $\widehat{Q}_{(-k)}$ plugged in, and $\lambda_{n,k}$ is a tuning parameter. Then, we estimate $w^* $ by
	\begin{equation}\label{eq:est_w}
	\widehat{ w}^{(k)} = \arg\min_{w} {E}_n^{(k)}\left[\nabla^2 l_\phi\left(\widehat{\beta}^{(k)};\widehat{\Omega}_+^{(-k)} ,\widehat{\Omega}_-^{(-k)}\right)\left(X_1- X_{-1}^\top w\right)^2\right] + \widetilde{\lambda}_{n,k}\|w\|_1,
	\end{equation}
	where $\widetilde{\lambda}_{n,k}$ is a tuning parameter. Let $\left(\widehat{\beta}_{\rm null}^{(k)}\right)^{\top}=\left (0,\left(\widehat{\beta}_{-1}^{(k)}\right)^{\top}\right)$, where $\widehat{\beta}_{-1}^{(k)}$ is a $p-1$ dimensional sub-vector of $\widehat{\beta}^{(k)}$ without $\widehat{\beta}_1^{(k)}$.  Finally, we construct the data-split de-correlated score test statistic $S^{(k)}(\widehat{\beta}_{\rm null}^{(k)}, \widehat{w}^{(k)})$ as  
	\begin{equation}\label{eq:split_score}
	S^{(k)}\left (\widehat{\beta}_{\rm null}^{(k)}, \widehat{w}^{(k)}\right ) = {E}_n^{(k)}\left[\nabla l_\phi\left(\widehat{\beta}^{(k)}_{\rm null};\widehat{\Omega}_+^{(-k)} ,\widehat{\Omega}_-^{(-k)}\right)\left(X_1-X_{-1}^\top\widehat{w}^{(k)}\right)\right].
	\end{equation}
	
	Combining $K$ data-split estimators, we can obtain the pooled estimator as
$
	\widehat{\beta}=K^{-1}\sum_{k=1}^K \widehat{\beta}^{(k)}.
$
        Likewise, the pooled de-correlated score test statistic is
$
	S = K^{-1}\sum_{k=1}^{K} S^{(k)}\left (\widehat{\beta}_{\rm null}^{(k)}, \widehat{w}^{(k)}\right). 
$
  
As shown in Theorem~\ref{thm:2}, under null hypothesis, we have
$
	n^{1/2}S\to N\left(0, \left(\nu^*\right)^\top \mathrm{var}\left[\nabla^2 l_{\phi}(\beta^*; \Omega_+, \Omega_-)XX^\top\right]\nu^*\right ).
$
The detailed algorithm is provided in Algorithm~\ref{algorithm:1}. In this algorithm, for a fixed $1\leq k\leq K$, $\widehat{\pi}_{(-k)}$ and $\widehat{Q}_{(-k)}$  are trained on a subset of samples of size $n(K-1)/{K}$. 
  
	\begin{minipage}{0.95\linewidth} 	 	  	
	\begin{algorithm}[H]\label{algorithm:1}
		\caption{Inference of $\beta^*$ using a sample-split procedure}
		\SetAlgoLined
		\KwIn{A random seed; $n$ samples; a positive integer $K$.}
		\KwOut{$\widehat{\beta}$ and a p-value for $\mathcal{H}_0:\beta^*_1=0$.}
		\label{algo1:1} Randomly split data into $K$ parts $\left\{{I}_k\right\}_{{k=1}}^K$ with equal size, and set $k=1$\;
		\label{algo1:2} Estimate $\pi$ and $Q$ on ${I}_{k}^c$ and denote the estimator as $\widehat{\pi}_{(-k)}$ and $\widehat{Q}_{(-k)}$\;
		\label{algo1:3} Obtain a data-split estimator $\widehat{\beta}^{(k)}$ on ${I}_k$ by~\eqref{eq:est_beta}, where ${\lambda}_{n,k}$ is tuned by cross-validation \;
		\label{algo1:4} Obtain an estimator $\widehat{w}^{(k)}$ for $ w^* $ by~\eqref{eq:est_w}, where $\widetilde{\lambda}_{n,k}$ is tuned by cross-validation \;
		\label{algo1:5}  Construct the data-split de-correlated score test statistic $S^{(k)}(\widehat{\beta}_{\rm null}^{(k)}, \widehat{w}^{(k)})$ by equation~\eqref{eq:split_score}, and the estimator of the variance $\widehat{\sigma}^2_k={E}_n^{(k)}\left[\left\{\nabla l_\phi\left(\widehat{\beta}^{(k)};\widehat{\Omega}_+^{(-k)} ,\widehat{\Omega}_-^{(-k)}\right)\right\}^2\left(X_1-X_{-1}^\top\widehat{w}^{(k)}\right)^2\right]$\;
		\label{algo1:6} Set $k=2,3,\ldots, K$, and repeat Step~2 and~5. Obtain $\left\{\widehat{\beta}^{(k)}\right\}_{k=1}^K$ and $\left\{S^{(k)}\left (\widehat{\beta}_{\rm null}^{(k)}, \widehat{w}^{(k)}\right )\right\}_{k=1}^K$ as well as $\left\{\widehat{\sigma}_k^2\right\}_{k=1}^K$.
		 Aggregate them by
		\begin{equation*}
			\widehat{\beta}=K^{-1}\sum_{k=1}^K\widehat{\beta}^{(k)},S = K^{-1}\sum_{k=1}^K S^{(k)}\left (\widehat{\beta}_{\rm null}^{(k)}, \widehat{w}^{(k)}\right ), \widehat{\sigma}^2 = K^{-1} \sum_{k=1}^K \widehat{\sigma}_k^2.
		\end{equation*}
		Calculate the p-value by $2\left(1-\Phi(n^{1/2}|S|/\widehat{\sigma})\right)$, where $\Phi(\cdot)$ is the cumulative distribution function of a standard normal distribution. 
	\end{algorithm}
	\end{minipage}
     	
	\subsection{Confidence intervals}
	
	We use the data-split de-correlated score to construct a valid confidence interval of $\beta^*$. This is motivated from the fact that the data-split de-correlated score $S^{(k)}\left(\beta, \widehat{ w}^{(k)}\right)$ is also an unbiased estimating equation for $\beta^*_1$ when fixing $\beta_{-1}=\beta^*_{-1}$. However, directly solving this estimating equation has several drawbacks,  such as the existence of multiple roots or ill-posed Hessian  (Chapter 5 in \citet{van2000asymptotic}). \citet{ning2017} proposed a one-step estimator, which solved a first order approximation of the de-correlated score. Following their procedure, we construct the data-split one-step estimator, $\widetilde{\beta}_1^{(k)}$, as the solution to, 
	\begin{equation*}
	S^{(k)}\left(\widehat{\beta}^{(k)}, \widehat{w}^{(k)}\right)+{E}_n^{(k)}\left[\nabla^2 l_{\phi}\left(\widehat{\beta}^{(k)};\widehat{\Omega}_+^{(-k)}, \widehat{\Omega}_-^{(-k)}\right)X_1(X_1- X_{-1}^\top\widehat{ w}^{(k)})\right](\beta_1-\widehat{\beta}^{(k)}_1)=0.
	\end{equation*}
	Hence, we have that
	$
	\widetilde{\beta}_1^{(k)}=\widehat{\beta}_1^{(k)}-S^{(k)}\left(\widehat{\beta}^{(k)}, \widehat{w}^{(k)}\right)/\widehat{I}_{1|-1}^{(k)},
	$
	where 
	\begin{equation*}
	\widehat{I}_{1|-1}^{(k)}={E}_n^{(k)}\left[\nabla^2 l_{\phi}\left(\widehat{\beta}^{(k)};\widehat{\Omega}_+^{(-k)}, \widehat{\Omega}_-^{(-k)}\right)X_1(X_1- X_{-1}^\top\widehat{w}^{(k)})\right].
	\end{equation*}
		
	Finally, the pooled one-step estimator is the aggregation of these data-split one-step estimators following
$
	\widetilde{\beta}_1=K^{-1}\sum_{k=1}^K\widetilde{\beta}_1^{(k)}.
$
	In Section~\ref{sec:theoretical}, we will show the asymptotic normality of the pooled one-step estimator $\widetilde{\beta}_1$, which provides a valid confidence interval for $\beta_1^*$. The algorithm for constructing confidence intervals is presented in Algorithm \ref{algorithm:2}. 
	
	\begin{minipage}{0.95\linewidth} 
		\begin{algorithm}[H]\label{algorithm:2}
		\caption{Confidence interval of $\beta^*_1$ using a sample-split procedure}
		\SetAlgoLined
		\KwIn{The data-split de-correlated score $S^{(k)}\left(\widehat{\beta}^{(k)}, \widehat{w}^{(k)}\right)$ and $\widehat{I}_{1|{-1}}^{(k)}$ for $k=1,\ldots, K$; $\widehat{\sigma}^2$ from Algorithm~\ref{algorithm:1}.}
		\KwOut{A 95\% confidence interval for $\beta^*_1$.}
		\label{algo2:1} Construct the data-split one-step estimator by 
		$
		\widetilde{\beta}_1^{(k)}=\widehat{\beta}_1^{(k)}-S^{(k)}\left(\widehat{\beta}^{(k)}, \widehat{w}^{(k)}\right)/\widehat{I}_{1|{-1}}^{(k)}
		$\;
		\label{algo2:2} Aggregate these data-split one-step estimators by
		$
		\widetilde{\beta}_1=K^{-1}\sum_{k=1}^K\widetilde{\beta}_1^{(k)},
		$
		and calculate
		$
		\widehat{I}_{1|{-1}} =K^{-1}\sum_{k=1}^K\widehat{I}_{1|{-1}}^{(k)}
		$\;
		\label{algo2:3} Construct the 95\% confidence interval by $\left(\widetilde{\beta}_1-1.96n^{-1/2}\widehat{\sigma}/\widehat{I}_{1|{-1}},\widetilde{\beta}_1+1.96n^{-1/2}\widehat{\sigma}/\widehat{I}_{1|{-1}} \right)$.
	\end{algorithm}
	\end{minipage}
	
	\subsection{Inference of the value}
	
	We adopt an analogy of the single-split procedure \citep{luedtke2016statistical, shi2020breaking} to infer the value under $D^*(X)$, $V(D^*)$, where $D^*(X)=\mathrm{sgn}\left(X^\top\beta^*\right)$. The single-split procedure splits the entire dataset into two parts. We use one part for training and nuisance parameter fitting, and conduct inference on the other part. When $\beta^*\propto\beta_{\rm opt}$, our procedure provides a valid inference procedure for the optimal value. The detailed procedure for inference of the value is presented in Algorithm  \ref{algorithm:3}.

	\begin{minipage}{0.95\linewidth} 
	\begin{algorithm}[H]\label{algorithm:3}
		\caption{Inference of the value $V(D^*)$ using a single-split procedure}
		\SetAlgoLined
		\KwIn{A random seed; $n$ samples.}
		\KwOut{A 95\% confidence interval for $V(D^*)$.}
		\label{algo3:1} Randomly split the data into two sets, $\widetilde{I}_1$ and $\widetilde{I}_2$ with sample size $n_1$ and $n_2$, and obtain $\widehat{\beta}$ using data in $\widetilde{I}_1$ by Algorithm~\ref{algorithm:1}\;
		\label{algo3:2} Estimate $\pi$ and $Q$ on $\widetilde{I}_1$ and denote the estimator as $\widehat{\pi}$ and $\widehat{Q}$\;
		\label{algo3:3} Estimate $V(D^*)$ on $\widehat{I}_2$ and denote the estimator as $\widehat{V}$,
		$
			\widehat{V}(\widehat{D})={{E}}^{(2)}_{n_2}\left[W_{\widehat{D}(X)}(Y,X,A,\widehat\pi,\widehat{Q})\right],
		$
		where $\widehat{D}(X)=\rm{sgn}(X^\top\widehat{\beta})$\;
		\label{algo3:4} Estimate the variance and denote the estimator as $\widehat{\sigma}^2_V$,
		$
			\widehat{\sigma}^2_V={\mathrm{var}}^{(2)}_{n_2}\left[W_{\widehat{D}(X)}(Y,X,A,\widehat\pi,\widehat{Q})\right],
		$
		where ${\mathrm{var}}^{(2)}_{n_2}(\cdot)$ is the sample variance on ${I}_2$.
		\label{algo3:4} Construct the 95\% confidence interval by $\left(\widehat{V}(\widehat{D})-1.96n_2^{-1/2}\widehat{\sigma}_V,\widehat{V}(\widehat{D})+1.96n_2^{-1/2}\widehat{\sigma}_V \right)$.
	\end{algorithm}
	\end{minipage}
	
	\section{Theoretical properties}
	\label{sec:theoretical}
	
	We assume the following conditions.   
	\begin{enumerate}[label=(C\arabic*)]
		\item \label{cond:c1} $\sup_{x\in \mathcal{X}}\|x\|_{\infty}$,  $\sup_{x\in \mathcal{X}}|x^\top {w^*}|$ and $\sup_{x\in \mathcal{X}}|x^\top \beta^*|$ are bounded by a sufficient large constant $\bar{c}$;
		$\sup_{x\in \mathcal{X}}|Q(a;x)|$ is bounded,
		and the conditional distribution of $Y(a)-Q(a;X)$ given $X$ is sub-exponential, i.e., it is either bounded or satisfies that there exists some constants $M,\nu_0\in {R}$ such that
		$$E\left[\exp\left\{|Y(a)-Q(a;X)|/M\right\}-1-|Y(a)-Q(a; X)|/M\mid X\right ]M^2\leq \nu_0/2,$$
		for both $a=1$ and $a=-1$.
		\item \label{cond:c7}  There exists some constants $0<\pi_{\min}<\pi_{\max}<1$ such that $\pi_{\min}\leq \pi(a;X)\leq \pi_{\max}$ with probability 1.
		\item \label{cond:c3} $\phi$ is convex and $\phi'(0)<0$; for any $t\in [-\bar{c}-\epsilon, \bar{c}+\epsilon]$ with some constant $\epsilon>0$ and a sequence $t_1$ satisfying $|t_1-t|=o(1)$, it holds that $0<\phi''(t)\leq C$ and $|\phi''(t_1)-\phi''(t)|\leq C|t_1-t|\phi''(t)$ for some constant $C>0$.
		\item \label{cond:c4} The smallest eigenvalue of $E[\nabla^2l_{\phi}(\beta^*; \Omega_+, \Omega_-)XX^\top ]$ is larger than $\kappa$, where $\kappa$ is a positive constant.
		\item \label{cond:c6} Suppose that for some $\alpha, \beta>0$, $\sup_{x}\left |\widehat{\pi}(a;x)-\pi(a;x)\right |=O_p(n^{-\alpha})$ and\\ $\sup_{x}\left |\widehat{Q}(a;x)-Q(a;x)\right |=O_p(n^{-\beta})$ for $a=1$ and $-1$, we require that $\alpha+\beta>1/2$. In addition, we require that 
		\begin{equation}\label{eqsparse1}
		\max\{s^*,s'\}\log p=o(n^{1/2}),
		\end{equation}
		and 
		\begin{equation}\label{eqsparse2}
		(n^{-\alpha}+n^{-\beta})s^*\to 0,
		\end{equation}
		 where $s^*=\|\beta^*\|_0$ and $s'=\| w^*\|_0$.
		\end{enumerate}
	Condition~\ref{cond:c1} on the joint distribution of $(X, A, Y)$ is commonly assumed in high-dimensional inference literature \citep{vandegeer2014,ning2017}. For  technical simplicity, we assume that the design is uniformly bounded. We also assume that $Y(a)-Q(a;X)$ is sub-exponential or bounded. This condition enables a faster convergence rate of high-dimensional empirical processes involving the estimation errors of $\widehat{\pi}$ and $\widehat{Q}$.  Under this condition,  if $\sup_{X}\left |\widehat{Q}(a;X)-Q(a;X)\right| = o_p(1)$, we have
	$$\left\|{E}_n\left [\left\{Y(a) -Q(a;X)\right\}\left\{Q(a; X)-\widehat{Q}(a;X)\right\}X\right]\right\|_{\infty}=o_p\left((\log p/n)^{1/2}\right).$$ Condition~\ref{cond:c7} prevents the extreme values in the true propensities.
	
	Condition~\ref{cond:c6} is imposed for Algorithm 1. We assume that it  holds on each split dataset. To simplify the notation, we do not distinguish $\widehat{\pi}$ and $\widehat{Q}$ with $\widehat{\pi}_{(-k)}$ and $\widehat{Q}_{(-k)}$ for a fixed $k$. 
First it requires that both $\widehat{\pi}$ and $\widehat{Q}$ are consistent and the convergence rates satisfy $n^{-\alpha-\beta}\ll n^{-1/2}$. This can be attained if either the convergence rate of $\widehat{\pi}$ or $\widehat{Q}$ is sufficiently fast.  For example, 
if $\pi$ is estimated by a regression spline estimator and is known to be  $p_{\pi}$-dimensional (low dimension) by design, we have $\sup_{X}\left|\widehat{\pi}(a;X)-\pi(a;X)\right|=O_p\left (n^{-1/3}\right )$, where $\pi$ is assumed to belong to the H$\ddot{\mbox{o}}$lder class with a smoothness parameter greater than $5p_{\pi}$ \citep{newey1997}. Then $n^{-\alpha-\beta}\ll n^{-1/2}$ is satisfied when $n^{-\beta}\ll n^{-1/6}$. Second, formula~(\ref{eqsparse1}) in Condition~\ref{cond:c6} requires that the number of nonzero entries of $\beta^*$ and $w^*$ is smaller than the order of $n^{1/2}/\log p$, which agrees with the conditions in the high-dimensional inference literature \citep{vandegeer2014,ning2017}. Finally, formula~(\ref{eqsparse2}) of Condition~\ref{cond:c6} 
indicates the convergence rates of the nuisance parameter estimations cannot be too slow if $s^*$ increases fast with the sample size $n$.

\begin{theorem}\label{thm:estimation}
        Assume that Conditions \ref{cond:c1}-\ref{cond:c6} hold. By choosing $\lambda_{n,k}\asymp (\log p/n)^{1/2}$ , we have
$
		\|\widehat{\beta}-\beta^*\|_1=O_p\left(s^* (\log p/n)^{1/2}\right).
$
\end{theorem}
Theorem~\ref{thm:estimation} assumes that both the outcome and propensity score models are correctly specified, $Q^m=Q$ and $\pi^m=\pi$ (implied by Condition~\ref{cond:c6}). Nonetheless, our proposed estimator enjoys the doubly robustness property in the sense that $\widehat{\beta}$ is still consistent if either $Q^m=Q$ or $\pi^m=\pi$. When $Q^m\not= Q$ and $\pi^m=\pi$, we have $\|\widehat{\beta}-\beta^*\|_1=O_p\left(s^*\max\left\{(\log p/n)^{1/2}, n^{-\alpha}\right\}\right)$; when $\pi^m\not= \pi$ and $Q^m=Q$, we have $\|\widehat{\beta}-\beta^*\|_1=O_p\left(s^*\max\left\{(\log p/n)^{1/2}, n^{-\beta}\right\}\right)$. This also indicates that as long as one of the estimators $\widehat \pi$ and $\widehat Q$ has a reasonably fast rate, the estimator $\widehat{\beta}$ is consistent. 

Theorems \ref{thm:2} and \ref{thm:3} provide the limiting distributions of the testing procedures in Algorithm~\ref{algorithm:1} and the pooled one-step estimator $\widetilde{\beta}_1$ in Algorithm~\ref{algorithm:2} via sample-splitting, respectively.


\begin{theorem}\label{thm:2}
		Assume that Conditions \ref{cond:c1}--\ref{cond:c6} hold. For Algorithm~\ref{algorithm:1}, under the null hypothesis $H_0: \beta_1^*=0$, by choosing $\lambda_{n,k}\asymp\widetilde{\lambda}_{n,k}\asymp  (\log p/n)^{1/2}$ , we have
$
		n^{1/2}S\to N(0,\sigma^2),
$
		and $\widehat{\sigma}^2 \to \sigma^2$, where $\widehat{\sigma}^2$ is given in Algorithm~\ref{algorithm:1}, and $\sigma^2=\left(\nu^*\right)^\top \mathrm{var}\left[\nabla^2 l_{\phi}(\beta^*; \Omega_+, \Omega_-)\right]\nu^*$.
	\end{theorem}


	\begin{theorem}\label{thm:3}
		Assume that Conditions \ref{cond:c1}-\ref{cond:c6} hold. The pooled one-step estimator satisfies that
		$$
			n^{1/2}\left(\widetilde{\beta}_1-\beta_1^*\right) {I}_{1|{-1}}^* \to N(0, \sigma^2),
		$$
		where $I_{1|{-1}}^*= {E}\left[\left\{\nabla^2 l_{\phi}(\beta^*;\Omega_+,\Omega_-)\right\}X_1\left(X_1-X_{-1}^\top w^*\right)\right]$. $\widehat{I}_{1|{-1}}$ is a consistent estimator for $I_{1|{-1}}^*$.
	\end{theorem}
 
\begin{remark}
Theorems~\ref{thm:2} and \ref{thm:3} assume that both the propensity and the outcome models are correctly specified and estimated. Nonetheless, when the propensity score is known by the design of the experiment, the conclusions in Theorems \ref{thm:2} and \ref{thm:3} still hold even if the outcome model is misspecified. In contrast, Q-learning requires correctly specified outcome models even when the propensity is known. 

In practice, an individualized treatment rule can still be linear even if the contrast function is non-linear. As such, our modeling framework is more flexible. The advantages of our methods extend to the high-dimensional setting. The outcome weighted learning approach does not involve modeling outcomes. However, the corresponding penalized estimator in the outcome weighted learning approach may have a slower convergence rate than the proposed estimator in Theorem \ref{thm:estimation} when the propensity score is estimated with a slow rate. Therefore, the de-correlated score or the one-step estimator based on the outcome weighted learning approach cannot achieve a limiting distribution with $n^{1/2}$ convergence rate as in Theorems~\ref{thm:2} and \ref{thm:3}. 
\end{remark}

To derive the asymptotic property of the inference procedure for the value, we further introduce the following conditions:
\begin{enumerate}[label=(C\arabic*)]
\setcounter{enumi}{5}
	\item \label{cond:c8}
		There exists an increasing function $\psi$ such that 1) $\psi(0)=0$; 2) there exists $\zeta>0$ and $\lim\sup_{t\to 0}\psi(t)/t^{\zeta}<+\infty$; 3) $|E(Y(1)-Y(-1)\mid X)|\leq \psi(|X^\top\beta^*|)$ when $|X^\top\beta^*|\leq t_0$, where $t_0$ is a constant.
	\item \label{cond:c9}
There exists constants $\gamma>0$ and $C_{\gamma}>0$ such that for any $t$ in some neighborhood of $0$, we have that
$
			\mathrm{pr}\left(0<\left |X^\top\beta^*\right|\leq t\right )\leq C_{\gamma}t^{\gamma}.
$
\end{enumerate}
	
	\begin{theorem}\label{thm:value_infer}
		Assume that $Y$ is bounded and denote the sample size of $\widetilde{I}_1$ as $n_1$ and $\widetilde{I}_2$ as $n_2$. In addition to the conditions in Theorem~\ref{thm:estimation}, we further assume $n_1^{-\alpha-\beta}n_2^{1/2}=o(1)$, Conditions~\ref{cond:c8} and~\ref{cond:c9} holds with $\left (s(\log p/n_1)^{1/2}\right )^{\zeta+\gamma}=o_p(n_2^{-1/2})$, then we have
$
			{n_2}^{1/2}\sigma_V^{-2}(\widehat{V}(\widehat{D})-V(D^*))\to N(0, 1),
$
		where 
		$$
			{\sigma}^{2}_V=\mathrm{var}\left[W_{{D}^*(X)}(Y,X,A,\pi,{Q})\right].
		$$
	\end{theorem}
Theorem~\ref{thm:value_infer} holds for both regular and non-regular cases. Condition~\ref{cond:c8} implicitly assumes that $\beta^*$ corresponds to the optimal individualized treatment rule. When Condition~\ref{cond:c8} fails, the inference of the value under $D^*(X)$ requires stronger assumptions (see Theorem~S5 in the supplementary materialf for details). In the simulation studies and application, we choose $n_1=n_2=n/2$. 
		
	\section{Simulation}
	\label{sec:simulation}
	
	In this section, we test our estimation and inference procedure under various simulation scenarios. Let $\Delta(X)=\left\{Q(1;X)-Q(-1;X)\right\}/2$ and $S(X)=\left\{Q(1;X)+Q(-1;X)\right\}/2$. We generate $X\sim N\left(0, I_{p\times p}\right)$, and $Y=A\Delta(X) + S(X)+\epsilon$,  $\epsilon\sim N(0, 1)$. Let $\beta^{\mathrm{opt}}=(1,1,-1,-1,0,\ldots, 0)^\top, \beta^*_S=(-1,-1,1,-1,0,\ldots, 0)^\top$, and $\beta_{\pi}^*=(1,-1,0,\ldots, 0)^\top$. The following scenarios are considered: (I) $\Delta(X)=\xi X^\top \beta^{\mathrm{opt}}$, $S( X)=0.4 X^\top \beta_S^*$, and $\pi(1; X)=\exp(0.4 X^\top \beta_{\pi}^*)/\left\{1+\exp(0.4 X^\top \beta_{\pi}^*)\right\}$; (II) $\Delta(X)=\left\{\Phi\left(\xi X^\top \beta^{\mathrm{opt}}\right)-0.5\right\}\times \widetilde{\Delta}(X)$, $S(X)=\exp\left(0.4 X^\top \beta_S^*\right)$, $\pi(1; X)=\exp\{0.25\times(X_1^2+X_2^2+X_1X_2)\}/\left[1+\exp\{0.25\times(X_1^2+X_2^2+X_1X_2)\}\right]$, where $\widetilde{\Delta}( X)=2(\sum_{l=1}^{4} X_l)^2+2\xi$ and $\Phi(\cdot)$ is the cdf of the standard normal distribution.

  Under these settings, the magnitude of the treatment effect $\Delta(X)$ changes with $\xi$, which ranges from $0.1$ to $1$. Scenario~(I) features a linear outcome model $Q(a;X)$ for both $a=1$ and $a=-1$, and a logistic model for the propensity.  
  Scenario~(II) has a nonlinear treatment effect $\Delta(X)$, though the decision boundary is still linear. The treatment assignment mechanism is also complex. 
More simulation results with a mixture of both discrete and continuous covariates, as well as highly correlated design matrices and non-regular cases, can be found in the supplemental materials.

We compare the pooled estimator with Q-learning, a regression-based method \citep{qian2011}.  With high-dimensional covariates, we fit a linear regression with a lasso penalty in Q-learning for all scenarios. The inference target  of interest is $\beta^{\mathrm{opt}}$. However, the limits of the coefficients estimates using either proposed method or Q-learning may not be identical to $\beta^{\mathrm{opt}}$. In our simulation experiments, we will test and construct confidence intervals for $\beta_l^*$'s, $ l=1,\ldots, 8$,  the $l$-th coordinate of $\beta^*$, which by abuse of notations, denote the limits of estimates under either method.  We generate large  data sets multiple times using the same data-generating process, and empirically verify that the sparsity pattern of $\beta^*$ matches with that of $\beta^{\mathrm{opt}}$. Hence, inferences on $\beta^*$  provide insights on the true optimal decisions.   
We conduct the hypothesis testing for Q-learning using the decorrelated score test proposed in  \citet{ning2017}, and construct 95\% confidence intervals for the coefficients of interest  in the context of Q-learning. For value inference, we implement the Algorithm~\ref{algorithm:3} as our proposed approach; for Q-learning, we implement the Algorithm~\ref{algorithm:3} with the coefficients $\widehat{\beta}$ estimated from Q-learning approach. The true value $V(\beta^*)$ is approximated by the average of estimated values on a large independent dataset. An R package called \texttt{ITRInference} is coded to
implement the proposed method and Q-learning approach. For the proposed method, the user can specify the method or select from a list of candidates to estimate nuisance parameters. In our implementation, we choose to estimate $\pi$ and $Q$ functions nonparametrically for all scenarios. To be more specific, we first implement a distance correlation-based variable screening procedure \citep{runze2012}. We then fit a kernel regression using the selected variables after screening. When estimating $\pi$, we set caps at $0.1$ and $0.9$ to trim extreme values.

	In all scenarios, the sample size $n$ and the dimension $p$ range from $350$, $500$, $800$, $1600$ to $2500$. We set the nominal significant level at $0.05$, and  the nominal coverage at $95\%$. We report the type I errors,  the powers of the hypothesis tests, and the value functions under the estimated decision rules out of 500 replications.  In particular, we present the type I errors for testing $\beta_5^*$ to $\beta_8^*$, and the powers for testing $\beta_1^*$ to $\beta_4^*$.    For each method, we also present the coverage of the interval estimations around the limiting coefficients.
	
\begin{figure}
		\centering
		\includegraphics[width=0.7\linewidth]{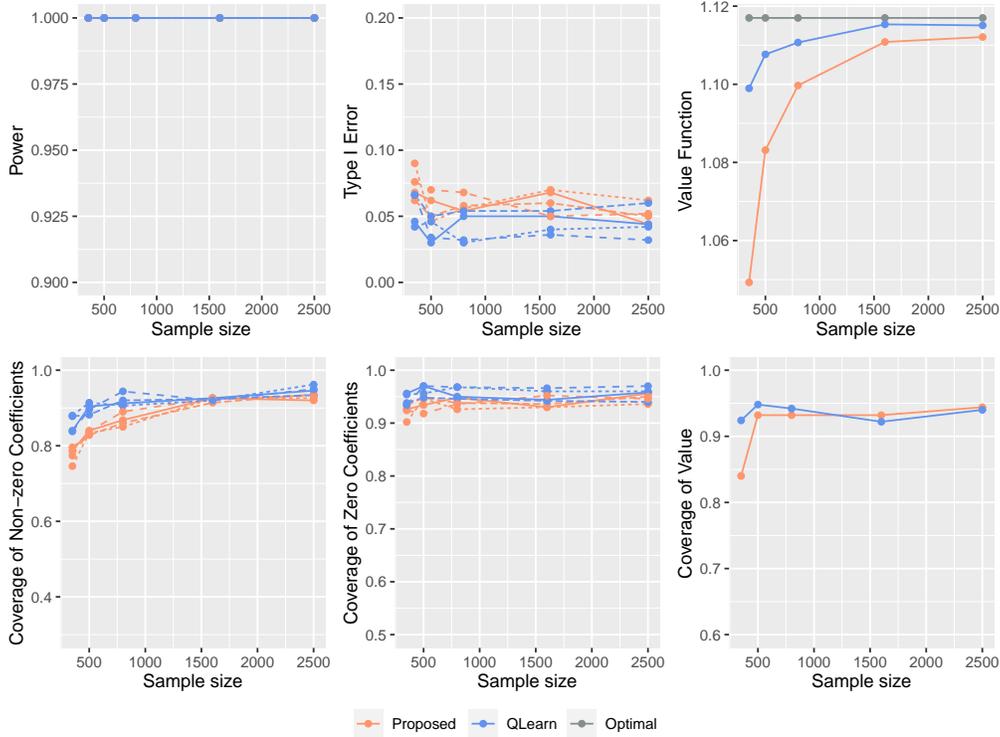}
		\caption{Simulation results for Scenario~(I) with the change of sample size when $\xi=0.7$ and $p=2500$. Types of the line represent different coefficients.}
		\label{fig:changesample1}
	\end{figure}

\begin{figure}
	\centering
	\includegraphics[width=0.7\linewidth]{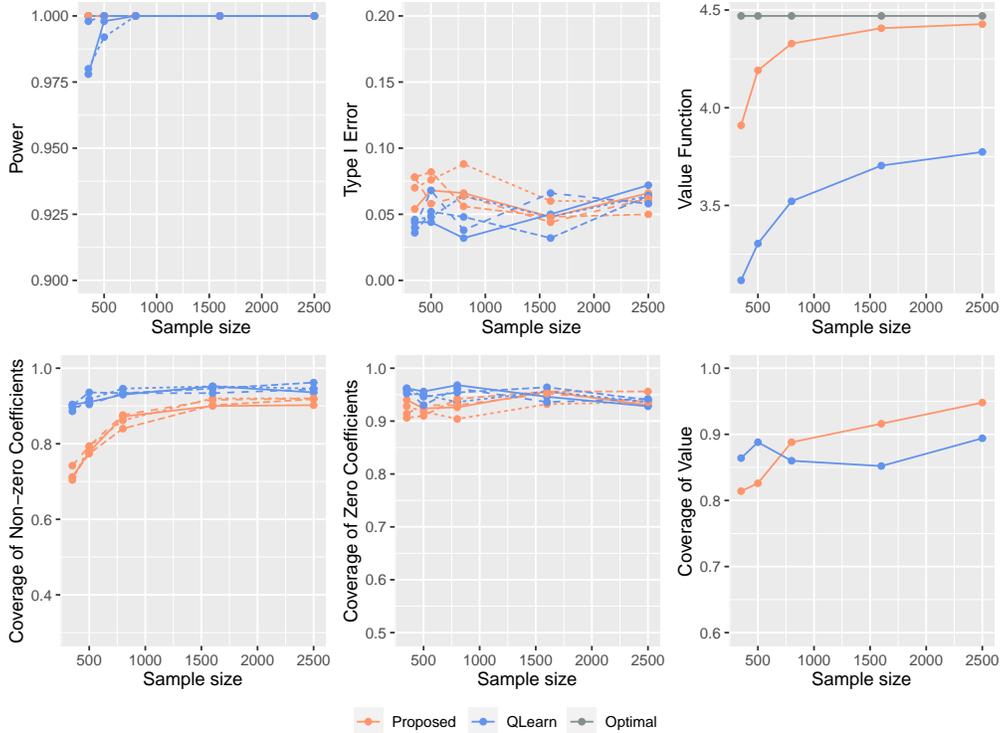}
	\caption{Simulation results for Scenario~(II) with the change of sample size when $\xi=0.8$ and $p=2500$. Types of the line represent different coefficients.}
	\label{fig:changesample3}
    \end{figure}

Figures~\ref{fig:changesample1} and~\ref{fig:changesample3} show the simulation results for different scenarios,  with the sample size $n$ varied and the $p$ and $\xi$ fixed. Additional results on varying $p$ with $n$ and $\xi$ fixed  can be found in the supplementary material. As expected, in Scenarios~(I) (Figure~\ref{fig:changesample1}) where the regression model is correctly specified for Q-learning, Q-learning yields a better value function.  Conversely,  the proposed method outperforms the Q-learning method in Scenario~(II) (Figure~\ref{fig:changesample3}). In terms of the type I error and power, the proposed method  is comparable  to the Q-learning approach in Scenario~(I) (Figure~\ref{fig:changesample1}). For Scenarios~(II) (Figure~\ref{fig:changesample3}), our method is more powerful, and the type I errors are well controlled. The power reduction for the Q-learning approach may be due to the model misspecification. The coverage of $\beta_5^*$ to $\beta_8^*$ are concentrated near $95\%$, and the coverage of the $\beta_1^*$ to $\beta_4^*$ gradually approach $95\%$ for the proposed method. For the coverage of the value $V(\beta^*)$, the inference procedure achieves a valid CI for the value under the proposed approach in both scenarios when the sample size is large enough. However, the inference for the value under the Q-learning is under coverage due to the model misspecification.


\section{Application to complex patients with type-II diabetes}
\label{sec:realdata}

In this section, we apply our proposed estimation and inference procedures to construct the optimal individualized treatment rule for complex patients with type-II diabetes. The data are collected from the electronic health records through Health Innovation Program at University of Wisconsin. The entire dataset includes $n=9101$ patients. There are 40  covariates, including socio-demographic variables, previous disease experiences, and baseline HbA1c levels, etc.  The outcome is the indicator whether the patient  successfully controls the HbA1c below $8\%$ after a year. The treatment  $A=1$ if the patient received any medications, including insulin, sulphnea or OHA,  and $A=-1$ otherwise.   Among $9101$ patients, $17.1\%$ had a missing post-treatment HbA1c measurement, and $15.4\%$ had the missing baseline HbA1c measurements. We impute missing values using  Multivariate Imputation by Chained Equations (\texttt{MICE} package in R), which is based on the estimated conditional distributions of each covariate given other covariates \citep{Stef2011}. To address the possible interactions among covariates, we consider both raw covariates and all first-order interactions. We rank these covariates by their variances and select $p=100$ covariates with top variances.

We split the dataset into a training dataset ($80\%$ of the entire dataset) and a testing dataset ($20\%$ of the entire dataset). The proposed method and Q-learning are fitted on the training dataset using the same strategies as described in simulation studies. 
To evaluate these estimated decision rules, we calculate the value function by
$
{E}_n [ {Y}{1 \{A=\widehat{D}(X) \}/\widehat{\pi}_0} ],
$
on the testing dataset, where $\widehat{D}$ is the estimated decision rules on the training dataset and $\widehat{\pi}_0$ is estimated the propensity scores on the testing dataset. 
The entire procedure is repeated $100$ times with random training and testing data splits. The mean and standard deviation (sd) of the value functions over these repeats are summarized in Table~\ref{tab:compare_realdata}. Both the proposed  and Q-learning methods construct decision rules that yield better results than the current clinical practice (sd of the difference is 0.0138 (Proposed); 0.0143 (Q-Learning)). Furthermore, our proposed method achieves a higher value function than Q-learning approach as shown in Table~\ref{tab:compare_realdata} (sd of the difference is 0.0115).  

Next, we conduct the inference procedure to identify driving factors of the optimal individualized treatment rule as well as to provide an interval estimation using the entire dataset. Results are presented in Table~\ref{tab:coef_realdata}. After controlling for the false discovery rate below $0.05$, our results indicate that a female patient with a higher HbA1c value at baseline are more likely to benefit from the treatment. The figure comparing the list of significant covariates selected by the proposed method and Q-learning can be found in the supplementary material (see page 11).

\begin{table}
	\caption{Results for comparisons on value functions. }
	\label{tab:compare_realdata}
	\begin{center}
		\begin{tabular}{lcc}
			\toprule
			Method & Mean &  Sd \\
			\hline  \\[-1.5ex]
	        Observed &0.860 &0.008 \\
	        PEARL&0.877 &0.015\\
	        Q-Learning & 0.869 &0.015\\
			\bottomrule
		\end{tabular}
	\end{center}
\end{table}


\begin{table}
	\caption{Coefficients and p-value for the identified significant covariate of the estimated optimal ITR. Special chronic conditions refer to chronic conditions including amputation, chronic blood loss, drug abuse, lymphoma, metastatistic cancer, and peptic ulcer disease.  Bucketized age refers to a variable created by bucketizing the raw age by its observed quartiles. }
	\label{tab:coef_realdata}
	{\footnotesize \begin{center}
		\begin{tabular}{lccc}
			\toprule
			\multicolumn{1}{c}{Covariate} & Coef &  P-value &95\% - CI\\
			\hline  \\
			Diabetes with Chronic Complications : Fluid and Electrolyte Disorders&-0.024&$4.71\times 10^{-2}$&[-0.047,-0.001]\\
			Diabetes with Chronic Complications : African American&-0.027&$3.58\times 10^{-2}$&[-0.052,-0.001]\\
			Alcohol Abuse : Entitlement Disability Indicator (Yes)&-0.054&$3.33\times 10^{-2}$&[-0.104,-0.004]\\
			HCC Community Score : Special Chronic Conditions&-0.022&$2.99\times 10^{-2}$&[-0.042,-0.002]\\
			Hypertension : Lower Extremity Ulcer&-0.036&$2.39\times 10^{-2}$&[-0.068,-0.005]\\
			HbA1c at Baseline : African American&0.019&$2.26\times 10^{-2}$&[0.003,0.036]\\
			Entitlement Disability Indicator (Yes) : Hypothyroidism&-0.024&$2.25\times 10^{-2}$&[-0.045,-0.003]\\
			Cardiac Heart Failure : Peripheral Vascular Disease&-0.029&$2.24\times 10^{-2}$&[-0.057, -0.001]\\
			Chronic Kidney Disease : HbA1c at Baseline&0.081&$1.97\times 10^{-2}$&[0.014, 0.149]\\
			Other Race (exclude White and Black) : Special Chronic Conditions&0.016&$1.95\times 10^{-2}$&[0.003,0.029]\\
			Liver Disease : Weight Loss&0.015&$1.72\times 10^{-2}$&[0.003,0.027]\\
			Other Neurological Disorders : Female&-0.021&$1.28\times 10^{-2}$&[-0.038,-0.005]\\
			Lower Extremity Ulcer : HbA1c at Baseline&0.039&$9.60\times 10^{-3}$&[0.010,0.069]\\
			Diabetes with Chronic Complications : Bucketized Age&0.040&$9.05\times 10^{-4}$&[0.016,0.063]\\
			HbA1c at Baseline : Female&0.044&$8.47\times 10^{-8}$&[0.028,0.061]\\	
			\bottomrule
		\end{tabular}
\end{center}}
\end{table}

\section{Discussion}
\label{sec:discussion}

In this paper, we consider a single stage problem and assume a high-dimensional linear decision rule. In practice, especially in managing chronic diseases, dynamic treatment regimes are widely adopted, where sequential decision rules for individual patients adapt overtime to the evolving disease. One future direction is to develop inferential methods in the multi-decision setup. We can also extend the linear decision rule to a single index decision rule $d(X^\top\beta^*)$, where $d$ is an unknown function. Throughout, we require that the surrogate loss function  be differentiable. A non-differentiable surrogate loss such as the hinge loss does not have a well-defined Hessian, which hinders the construction of the de-correlated score. This can be addressed by a smoothed hinge loss or an approximation of the Hessian. We are currently working on these possible extensions.

In this work, we adopt the de-correlated score to infer the high-dimensional linear decision rule. It is also possible to utilize other high-dimensional influential tools developed recently. Partial penalized tests proposed in \cite{shi2019linear} allows to test hypothesis involving a growing number of coefficients as sample size increases. \cite{ma2020global} consider the global and simultaneous hypothesis testing for high-dimensional logistic regression models. Although a modified algorithm~\ref{algorithm:1} can be combined with these methods, its theoretical property, especially the consequences of nuisance parameter estimation with slow rates, need future investigations.
 
Another future work is to extend the proposed approach to multiple treatment options setup.  There are several possible directions. First direction is to transform the multiple treatment problem to multiple binary decision problems. We can consider a sequential decision making strategy \citep{zhou2018outcome} by conducting a series of binary treatment selections.  It is shown that such strategy is Fisher consistent.  Another direction is to adopt techniques used in multi-label classification problem to estimate the optimal individualized treatment rule \citep{liang2018estimating}. We can incorporate the weights based on outcome model and propensity model into this framework and develop the corresponding inferential procedures. We are currently working on these extensions.
	
	\newpage
	
	\begin{center}
		{\large\bf SUPPLEMENTARY MATERIALS}
	\end{center}
	
	\begin{description}
		
		\item[Supplemental Materials:] The supplementary material includes additional simulation settings and proofs of lemmas and theorems. The R package called \texttt{ITRInference} is available at  \href{https://github.com/muxuanliang/ITRInference}{ITRInference Package}.
		
	\end{description}

\singlespacing
\bibliographystyle{apalike}
\bibliography{ref.bib}

\end{document}